\documentclass[12pt,preprint]{aastex}
\shorttitle{Magnetic field morphology of the white dwarf WD\,1953-011}
\shortauthors{Valyavin et al.}
\begin{document}

\title{The pecular magnetic field morphology of the white dwarf WD\,1953-011:
evidence for a large-scale magnetic flux tube?}

\author{G. Valyavin\altaffilmark{1}, G.A.~Wade\altaffilmark{2},
S.~Bagnulo\altaffilmark{3}, T.~Szeifert\altaffilmark{4},
J.D.~Landstreet\altaffilmark{5}, Inwoo Han\altaffilmark{1},
A.~Burenkov\altaffilmark{6}}

\altaffiltext{1}{Korea Astronomy and Space Science Institute, 61-1,
Whaam-Dong, Youseong-Gu, Taejeon, Republic of Korea 305-348 }
\altaffiltext{2}{Physics Department, Royal
Military College of Canada, Kingston, Ontario, Canada}
\altaffiltext{3}{Armagh Observatory, Northern Irland}
\altaffiltext{4}{European Southern Observatory, Alonso de
C\'ordova 3107, Santiago, Chile}
\altaffiltext{5}{Physics \& Astronomy Department, University of Western
Ontario, London, Canada}
\altaffiltext{6}{Special Astrophysical Observatory, Russian Academy
of Sciences, Nizhnii Arkhyz, Karachai Cherkess Republic, 357147,
Russia}

\begin{abstract}
We present and interpret new spectropolarimetric
observations of the magnetic white dwarf WD\,1953-011.
Circular polarization and intensity spectra of the H$\alpha$ spectral
line demonstrate the presence of two-component magnetic field
in the photosphere of this star. The geometry consists of a weak, large
scale component, and a strong, localized component. Analyzing
the rotationally modulated low-field component, we establish a rotation
period $P_{rot} =  1.4480 \pm 0.0001$~days.
Modeling the measured
magnetic observables, we find that the low-field component
can be described by the superposition of a dipole and quadrupole.
According to the best-fit model, the inclination of the stellar rotation axis with
respect to the line of sight is $i \approx 20^\circ$, and the angle between
the rotation axis and the dipolar axis is $\beta \approx 10^\circ$.
The dipole strength at the pole is about 180\,kG, and the
quadrupolar strength is about 230\,kG. These data suggest a
fossil origin of the low-field component. In contrast,
the strong-field component exhibits a peculiar, localized structure
(``magnetic spot'') that confirms the conclusions of Maxted and co-workers.
The mean field modulus of the spot ($|B_{spot}| = 520 \pm 7$~kG)
together with its variable longitudinal magnetic field
having a maximum of about $+400$~kG make it difficult to
describe it naturally as a high-order component of the star's global poloidal
field. Instead, we suggest that the
observed strong-field region has a geometry similar to a magnetic flux
tube.
\end{abstract}

\section{Introduction}
\label{Intro}

At present, there are more than one hundred known isolated
magnetic white dwarfs (MWDs) with
magnetic field strengths from a few tens of kilogauss to several hundreds of
megagauss  \citep{ABL81,SS95,LBH03,VBFM03,AZC04,VBF06,KAW07,JAN07}. It is generally
assumed that the
magnetic fields of the strong-magnetic MWDs (those with MG-strength fields)
are organized as
low-order multipolar fields with dominating dipolar components
\citep{P99}.
The rotation periods and surface magnetic fields of the strong-magnetic
MWDs are believed to be stable on long time scales
\citep{SN91}, suggesting that their
fields are fossil remnants of the fields of their progenitor stars.
A comparison of the field strengths and incidence statistics of the strong-magnetic
MWDs with magnetic fields of Ap/Bp main sequence stars support this
assumption \citep{ABL81}.

Despite the progress with the strong-magnetic MWDs, the
magnetic properties of the weak-field degenerates are only poorly known.
Presently, only a few white dwarfs with kilogauss magnetic fields has been
identified \citep{SS95,FVB03,VBFM03,AZC04,VBF06,KAW07,JAN07}. Their rotation and field
geometries are poorly studied, although some progress has been achieved
by \citet{MFMW00} and \citet{Wad03}
with a study of WD\,1953-011 and by \citet{VBMF05} with WD\,0009+501.
\citet{MFMW00} established that the magnetic morphology of
WD\,1953-011 can be described by both low-field  ($B \sim 90$\,kG) and
strong-field ($B \sim 500$\,kG) components. Some evidence for the presence
of a non-dipolar (quadrupolar) component
was also found in WD\,0009+501 by \citet{VBMF05}.

Motivated by the results of \citet{MFMW00}, we have undertaken
collaborative spectropolarimetric monitoring of WD1953-011. In this paper
we report results of these observations and analyse them in the manner
presented by \citet{Wad03} and \citet{VBMF05}.
Our goal is to determine precisely the magnetic morphology of this degenerate.

\section{A few preliminary remarks}
\label{Rem}

\begin{figure}
\centering
\includegraphics[width=18.5cm, height=14cm, angle=0]{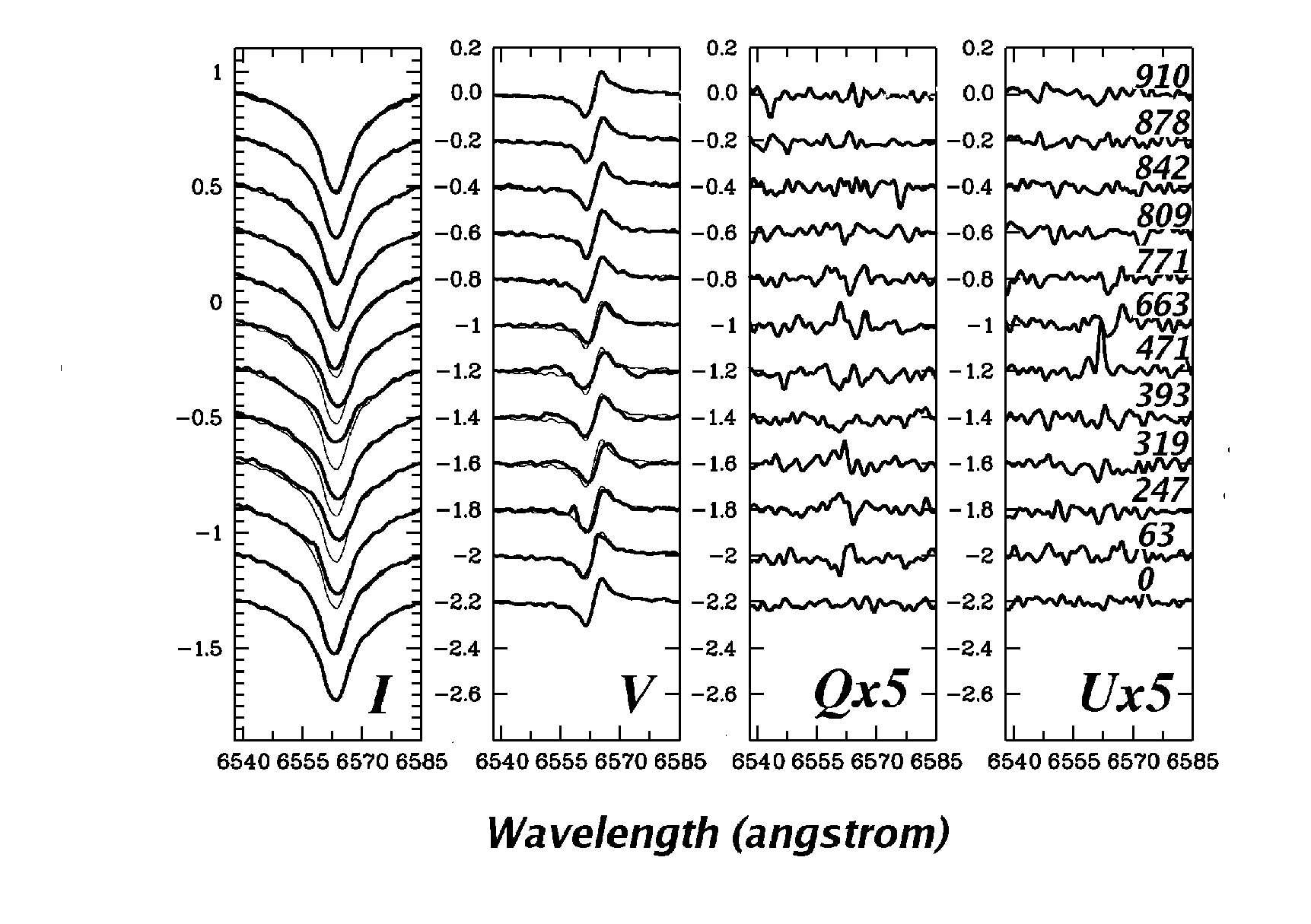}
\caption{Stokes $IQUV$  H$\alpha$ profile timeseries of WD1953-011,
obtained using the FORS1 spectropolarimeter at the ESO VLT. Phases
correspond to the magnetic ephemeris obtained in this paper
and are expressed in part per mil at right. The thin lines
represent the observations obtained at phase 0, and are reproduced
to emphasise the variability of the Stokes profiles.
}
\label{fig1}
\end{figure}

Our preliminary analysis of the spectropolarimetric data obtained with FORS1
at the VLT \citep{Wad03} revealed significant variability of the
Stokes~$I$ and $V$ spectra of WD\,1953-011 due to rotation, with a period of about
1.45 days. In Fig.~\ref{fig1} we show those results which illustrate the
variation
of the Stokes~$I,V,Q$ and $U$ profiles of the H$\alpha$ line with rotational
phase (phase increases from top to bottom in the figure).
As can be seen, the Stokes~$I$ profile is strongly variable. The central S-wave of
the Stokes~$V$ profile is almost constant. However, near rotational
phase 0.6 additional broadened Stokes~$V$ signatures appear in the
H$\alpha$ wings. These signatures correspond to the weak ``satellite
features'' observed by \citet{MFMW00} in the wings of
the H$\alpha$ profile at these phases, and were attributed to
the presence of a high-field magnetic structure. Linear polarization
Stokes~$Q$ and $U$ signatures are only marginally detected at several
rotational phases.

In this paper we extend this analysis using addition observational
material obtained with the AAT \citep{MFMW00} and with the
6-m Russian telescope BTA, and using more sophisticated modeling techniques.

\section{Observations}
\label{Obs}

Spectropolarimetric observations of WD~1953$-$011 were obtained in
service mode between May and June 2001 with FORS1 on the ESO
VLT. FORS1 is a multi-mode instrument for imaging and multi-object
spectroscopy equipped with polarimetric optics, and is described by
\citet{AFF98}. For this work, FORS1 was used to measure
Stokes~\textit{IQUV} profiles of WD~1953$-$011 at 12 different
rotation phases, using grism 600\,R (plus order separation filter
GG\,435), which covers the interval 5250\,\AA -- 7450\,\AA. With
a slit width of 0.7", the spectral resolving power was about 1650.
To perform circular polarization measurements, a $\lambda/4$ retarder
waveplate and a Wollaston prism are inserted in the FORS1 optical path
(see \citet{A67}). The $\lambda/4$ retarder waveplate can be
rotated in 45\degr\ steps.

To perform linear polarization
measurements, a $\lambda/2$ retarder waveplate is used, which can be
rotated in $22.5^\circ$ steps.  At each epoch, Stokes~$V$ was measured
taking two 420\,s exposures: one with the $\lambda/4$ retarder
waveplate at $-45^\circ$, and one with the $\lambda/4$ retarder
waveplate at $+45^\circ$. Stokes~$Q$ and $U$ were measured taking four
600\,s exposures with the $\lambda/2$ retarder waveplate at $0^\circ$,
$22.5^\circ$, $45^\circ$, and $67.5^\circ$. The Stokes $V/I$
circular polarization spectrum was then obtained by calculating
\begin{equation}
P_V = \frac{V}{I} = \frac{1}{2} \left(r_{-45} - r_{+45}\right)
\end{equation}
where
\begin{equation}
\label{erre}
r_\alpha = \frac{f^{\rm o} -f^{\rm e}}{f^{\rm o} +f^{\rm e}}.
\end{equation}
In Eq.(2) $f^{\rm o}$ is the flux measured in the ordinary beam and
$f^{\rm e }$
is the flux measured in the extra-ordinary beam, obtained with the
$\lambda/4$ retarder waveplate at angle $\alpha$. Similarly, the linear
polarization was obtained by calculating
\begin{equation}
\begin{array}{rcl}
P_Q = \frac{Q}{I} &=& \frac{1}{2} \left(r_{0} - r_{45}\right)        \\
P_U = \frac{U}{I} &=& \frac{1}{2} \left(r_{22.5} - r_{67.5}\right) \\
\end{array}
\end{equation}
where $r_\beta$ is defined by Eq.~\ref{erre}, except that $\beta$
refers to the position angle of the $\lambda/2$ retarder
waveplate. Fluxes $f^{\rm o}$ and $f^{\rm e}$ were obtained from the
raw data after bias correction and wavelength calibrations performed
using standard IRAF routines.

\begin{figure}
\centering
\hspace*{-0.5cm}
\includegraphics[width=16cm, height=19cm, angle=270]{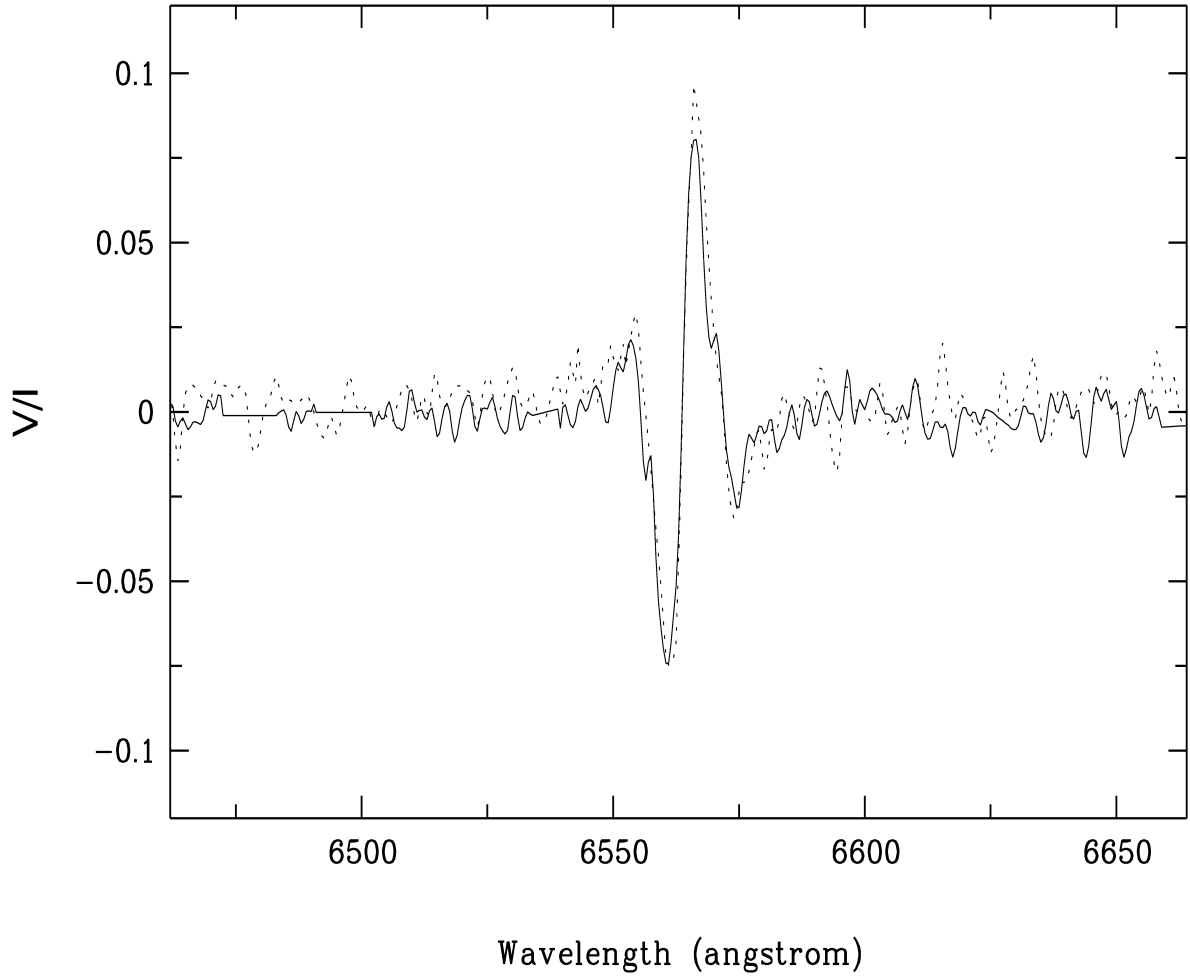}
\caption{Stokes $V$ at the H$\alpha$ line obtained with the VLT
(the solid line) and the BTA (the dashed line). The spectra correspond
to phases of the maximum visible strong-field Zeeman satellite features
at the H$\alpha$ wings ($\phi \approx 0.3$ in observations with the BTA and
$\phi \approx 0.5$ in observations with the VLT: the observed phase shift is
discussed in Sect.\,8 and Sect.\,9 of this study.)
}
\label{fig2}
\end{figure}

These observations were supported by a short
observing run at the 6-m Russian telescope BTA where we obtained
additional $I, V$ series of spectra of WD\,1953-011 using the UAGS
spectropolarimeter, with nearly the same resolving power as
in the observations with FORS1. The instrument is described in detail
by \citet{ABVD95} and by \citet{NVF02}. The
observational technique and data reduction are similar to
those described by \citet{Bag00,Bag02} and by
\citet{VBMF05}. A comparative analysis of the spectropolarimetric data
obtained from the different telescopes showed identical results. A comparison
of the Stokes~$V$ spectra obtained with the VLT and BTA is illustrated in
Fig.~\ref{fig2}. The spectra are obtained at different times but
similar rotational phases.

In addition to the spectropolarimetric data from the VLT and BTA, in
this paper we also use high-resolution spectroscopic data (Stokes~$I$)
obtained at the AAT and described by \citet{MFMW00}. Together
with the spectropolarimetry, these data extend the analysis presented by
\citet{Wad03} to a much longer time base.
Table~\ref{tbl1} gives an overview of all the observations. In
the table: {\bf JD} is the Julian Date; {\bf Exp} is an equivalent exposure time of an
observation; {\bf Stokes} is the observed Stokes parameter ($I,V,Q$ or $U$);
and {\bf Telescope} is telescope used (VLT, AAT or BTA).

\section{Mean field modulus of the magnetic field of
WD\,1953-011}
\label{Meanf}

We begin with an analysis of the low and high-resolution Stokes~$I$ spectra,
extending over 5 years. These spectra are used for
establishing the rotation period of the star,
as well as the
mean field modulus of the low- and strong-field components.

\subsection{The low-field component}

The low-field component of WD\,1953-011 was first discovered
spectroscopically by \citet{KDWA98} and described
in detail by \citet{MFMW00}. The mean field modulus,
$|B_G|$\footnote{For distinctness we label all magnetic observables related
to the low-field component with the subscript ``$G$'', assuming its large scale (Global
) geometry. Observables related to the strong-field component will be labeled
with the subscript ``$Spot$'' or ``$S$''.} exhibits a low-amplitude variation due to the
star's rotation, with a period estimated between hours and days
\citep{MFMW00}. These conclusions were made on the basis of the
high-resolution spectroscopy of the Zeeman pattern in the H$\alpha$ core.

\begin{figure}
\centering
%\hspace*{-1.cm}
\includegraphics[width=6.8cm, height=10.0cm, angle=270]{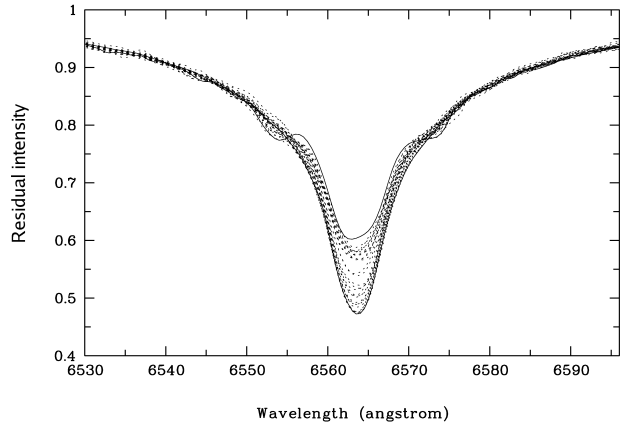}
\caption{The H$\alpha$ profiles obtained at the VLT, AAT and BTA.
High resolution spectra are convolved to the spectral resolution of
the FORSE1 and UAGS. The solid lines illustrate profiles at two extreme
rotation phases at which the spot component is most clearly seen
(the shallowest profile) and where the spot component is
absent (the deepest profile).
}
\label{fig3}
\end{figure}

In our low-resolution FORS1 and BTA observations, Zeeman splitting attributed to
the low-field component cannot be resolved spectroscopically.
In the spectra the splitting is revealed as
an additional variable broadening and desaturation of the H$\alpha$ core.
In this case, measurements of the low-field component
can be carried out by an analysis of the equivalent widths ($EW_{core}$)
of the H$\alpha$ core. Using field modulus $|B_G|$ values determined
by \citet{MFMW00} from an analysis of individual high-resolution H$\alpha$
line profiles and measuring equivalent widths of the H$\alpha$ cores, we may try
to calibrate the relationship $EW_{core}$\,--\,$|B_G|$ to allow us to
determine $|B_G|$ in the low-resolution spectra. To obtain the required calibration, we
estimated equivalent widths of the H$\alpha$ core
from the high resolution spectra obtained by \citet{MFMW00}.
In order to work with measurements having a uniform resolution, all high-resolution spectra were convolved
with a gaussian instrumental profile to reproduce the spectral
resolution of FORS1 and UAGS. The resultant spectra are presented in
Fig.~\ref{fig3}.

As one can see in Fig.~\ref{fig3}, the profiles are
strongly variable. The central intensity of the core also correlates with the
intensity of the strong-field Zeeman features which are found in the wings of
the H$\alpha$ profile (Fig.~\ref{fig3}: the two satellite features at
$\pm$10~\AA \, around the H$\alpha$ core). This correlation (the higher the
intensity of the features, the weaker the central intensity) is due to the
fact that the spot, which appears periodically on the visible disc due to rotation,
redistributes the flux according its projected area.
It is also seen that the width of the H$\alpha$ core
is variable itself due to the variable Zeeman pattern of the line core.
Therefore, to minimize the influence of the variable high-field spectral
features in measurements of the central Zeeman pattern attributed to the
low-field component, we artificially re-normalized all the profiles to equal
residual intensities ($r_c = 0.47$ at the line center) and measured equivalent widths of
the central narrow portion ($r_c \le 0.6$) of the resultant H$\alpha$
profiles. In these conditions, the variation of the $H\alpha$ core is
attributed only to the rotationally modulated low-field component. As hoped, we find a close correlation between the $EW_{core}$ measured in this way and the value of $|B_G|$ measured by Maxted. This relation is shown in Fig.\,4. 

Finally, using the $|B_G|-EW$ relationship derived from the
high-resolution spectra as illustrated in Fig.\,4, we inferred the field
modulus $|B_G|$ associated with each of the low-resolution spectra
(see Table~\ref{tbl2}).

\subsection{The strong-field component}

In order to measure the magnetic field modulus $|B_{Spot}|$ of the
strong-field component we
deblended the H$\alpha$ profile by means of a simultaneous fit of five
Gaussian profiles (three central profiles used to fit the H$\alpha$ core,
and two satellite gaussians to reproduce the strong-field Zeeman pattern).
This method enabled us to reproduce the Zeeman splitting of the strong-field
component and the corresponding magnetic field strength in those
spectra where the strong-field spectral features are seen.
The method also allows us to estimate the projected fractional area $S$ of the strong-field
area on the disc. Reconstructing by gaussians and extracting the Zeeman pattern
of the strong field component from the observed H$\alpha$ line profiles,
we determined $S$, the fraction of the flux absorbed by the strong-field pattern
relative to the total H$\alpha$ absorption. The method is rather rough and can be
considered as a first-guess approximation that is necessary for
the analysis described below. A more realistic calculation of the size of the
strong-field area is performed in Sect.\,8 where we model the spectra.
The results ($|B_{spot}|$ and $S$) are presented
in Table~\ref{tbl3}. $S$ is given in per cent of the disc area.

\section{
Mean longitudinal magnetic field of WD1953-011}

From the Stokes~$I$ and $V$ spectra obtained with the VLT and BTA we
determined longitudinal fields through the weak-field approximation
\citep{AMS73} modified to the analysis of the two-component
Stokes~$V$ spectra:

\begin{equation}
V(\lambda) \sim
(1 - S) B_G^l \, \bigg{(} \frac{\lambda}{\lambda_0} \bigg{)}^2
\frac{1}{I(\lambda)} \frac{{\rm d} I(\lambda) }{{\rm d} \lambda} \,\,
+ S V(\lambda)_{Spot}
\,
\end{equation}
where $B_G^l$ is the longitudinal field of the low-field component,
$\frac{d I(\lambda) }{d \lambda}$ describes the gradient of the flux
profile, $S$ is the relative area of the spot projected on the disc,
$\lambda_0$ is the H$\alpha$ rest wavelength and $V(\lambda)_{Spot}$ is
the Stokes~$V$ profile from the strong-field component observed in the
H$\alpha$ wings.

In Eq.(4) we effectively separate the disc
into two equivalent areas with different averaged magnetic field strengths.
The first term in the equation describes the weak-field area and the second
term is attributed to the strong-field component. The first
term is used in the usual manner according to which
the flux and its gradient are taken directly from the observed spectra.
(Inaccuracies due to the presence of the strong-field features in the wings
are comparatively weak: these features are
located quite far from the line core and do not affect the narrow
central low-field polarization profile). However, circular polarization
$V(\lambda)_{Spot}$ from the strong-field component (the second term in the equation)
cannot be fitted in the same way.

In order to fit $V(\lambda)_{Spot}$ and estimate the
longitudinal field of the strong-field area, we compute:

\begin{equation}
V(\lambda)_{Spot} = \frac{I(\lambda)^{L}_{Spot} - I(\lambda)^{R}_{Spot}}{I(\lambda)}
\end{equation}
\\

\noindent
where the flux $I(\lambda)$ is the observed H$\alpha$ flux profile, and
$I(\lambda)^{L}_{Spot}$ and $I(\lambda)^{R}_{Spot}$ are the left- and right-hand
polarized parts of the H$\alpha$ profile from the strong-field
equivalent area of the disc.
In the observed polarization spectra
$I(\lambda)^{L}_{Spot}$ and $I(\lambda)^{R}_{Spot}$ are mixed with fluxes
from the weak-field equivalent area and therefore can not be extracted
directly. However, we may estimate them with some simplifications.

Individually, $I(\lambda)^{L}_{Spot}$ and $I(\lambda)^{R}_{Spot}$ are
Zeeman-split profiles of the
circularly polarized satellite $\sigma$ components.
Due to the fact that the $\sigma_-$ component
is absent in $I(\lambda)^{R}_{Spot}$, and the $\sigma_+$ component is absent in
$I(\lambda)^{L}_{Spot}$, their centers of gravity are displaced, indicating the
presence of the longitudinal field from the strong-field area. Their
difference provides the non-zero circular polarization (Eqn.\,5).

Because $I(\lambda)^{L}_{Spot}$ and
$I(\lambda)^{R}_{Spot}$ are not resolved in the total left- and
right-circularly polarized observed fluxes, the
determination of their true shapes requires
detailed modeling the field geometry. However,
as a first-guess approximation we may describe them by simulating an
equivalent ``mean'' Zeeman-broadened H$\alpha$
profile magnetically displaced to the left- and right-sides from the rest
wavelength
 \footnote{The low resolving power of the FORS1
and UAGS, as well as unresolved circular polarization features
attributed to the strong-field area, enable us to consider the problem in
terms of Zeeman broadening instead of detailed analysis of the strong-field
Zeeman pattern.}. With this simplification only two
parameters~-- Zeeman broadening
and their magnetic displacement due to the
averaged longitudinal field from the spot should be varied to reproduce the
observed circular polarization.

In the simulation procedure we may use a template H$\alpha$
profile, artificially broadened to unresolved Zeeman patterns
typical for $I(\lambda)^{R}_{Spot}$ and $I(\lambda)^{L}_{Spot}$ and
magnetically displaced. This template can be
taken from a zero magnetic field solution for the atmosphere of
WD\,1953-011 or from the observed spectra.
For example, assuming the pressure-temperature
conditions in the spot area to be similar to conditions in the other parts
of the white dwarf's surface we may choose as the template one of
the observed weak-field H$\alpha$ profiles (obtained at those moments
when the spot is not seen). In our analysis we proceed this way.

Thus, to simulate $I(\lambda)^{L}_{Spot}$ and $I(\lambda)^{R}_{Spot}$
in order to fit the circular polarization (5) from the strong-field area and
estimate its longitudinal magnetic field we used the following iterative
method: \\

\begin{itemize}

\item
{\bf Step-1:} We construct the reference weak-field ``template'' H$\alpha$ profile
from the observed I-profiles obtained at those rotational phases where
the strong-field Zeeman pattern is not seen.\\

\item
{\bf Step-2:} We artificially broaden the template profile by a gaussian
filter with an arbitrary half-width
to an unresolved strong-field Zeeman pattern and displace the result
by the magnetic displacement factor $\Delta \lambda$ to the shorter /  longer wavelengths
to estimate the
$I(\lambda)^{L}_{Spot}$ and $I(\lambda)^{R}_{Spot}$ profiles.

\item
{\bf Step-3:} Varying the magnetic broadening of the estimated profiles
$I(\lambda)^{L}_{Spot}$ and $I(\lambda)^{R}_{Spot}$ and their Zeeman
displacement we finally fit the strong-field circular polarization (5) in
the working equation (4). The displacement found
$\Delta \lambda = 4.67 \cdot 10^{-13} B_S^l \lambda_0^2$ \citep{L80},
gives an estimate of the longitudinal field $B_S^l$.
(In other words, taking $S$ measured from the Stokes~$I$ spectra and presented
in Table~\ref{tbl3} we simultaneously fit the observed
combined Stokes~$V$ (4) varying $B_G^l$ and circular polarization (5) of the
strong-field component, where $B_S^l$ is one of the parameters.)
\end{itemize}

In the fit procedure, the associated error bars are obtained
using the Monte Carlo modeling method presented by
\citet{SS94}. An example of the fit is presented in Fig.~\ref{fig5}.
The results are collected in Table~\ref{tbl4}.

This method gives quite robust estimates of the longitudinal magnetic
field of the low-field component. In the case of the strong-field component,
the real intensities of the fields could be slightly over or underestimated due
to the simplifications described above.
For these reasons, estimates of the strong-field
component given here could be considered to be approximate.
As an alternative, the two-component circular polarization spectra
could be analyzed by using Zeeman tomography
\citep[for instance]{EJB02}. To provide 
more precise modeling,
below (Sec.\,8) we analyze our data again in the framework
of simplified Zeeman tomography.

\section{Period determination}
\label{Temporal}

To search for the star's rotation period we used the
equivalent widths $EW_{core}$ of the H$\alpha$
core determined in Sect.\,4 . This observable is the most sensitive indicator
for the determination of the rotation period.
To determine the rotation period we applied the Lafler-Kinman
method \citep{LK65}, as modified by
\citet{G04}. Analysis of the power spectrum of the data revealed a
signal indicating a probable period between 1.4 and 1.5~days
(Fig.~\ref{fig6}). This is consistent with the period estimate
($P \approx 1.45$\,days) given by \citet{Wad03} and
\citet{BM05}.

Detailed study
of the periodogram showed that the most significant sinusoidal signal
corresponds to a period P~=~1.4480~$\pm$~0.0001~days.  Other peaks are
located around 1.447 days and 1.442 days.
An examination of these periods reveals distorted,
non-sinusoidal
signals and we do not consider these periods further.

The phase variation of the H$\alpha$ core equivalent widths $EW_{core}$ derived
with this period is presented in Fig.~\ref{fig7}.
The derived period shows a very good agreement among all the observations
taken from different telescopes (the VLT, BTA and AAT). For the minimum of
$EW_{core}$ we obtain the following ephemeris:
\[
\mathrm{JD} = 2452048.801 \pm 0.03 + 1^d _\cdot 4480 \pm 0.0001\,\mathrm{E}
\]

The corresponding phase curves of the mean field modulus $|B_G|$
and longitudinal field $|B_G^l|$ of the weak-field components
are presented in Fig.~\ref{fig8}. The phase curves
are almost
sinusoidal, and symmetric about the values of about $|B_G|$~=~$+87$\,kG
and $|B_G^l|$~=~$-43$\,kG.  The
modulus of the weak-field component varies from $+77 \pm 1.5$\,kG to $+97 \pm
1.5$\,kG; the longitudinal magnetic field shows variation in the
range $-39 \pm 2$\,kG to $-47 \pm 2$\,kG.

As one can see in Fig.~\ref{fig8} the behavior of the longitudinal
magnetic field and modulus of the weak-field component suggest the field
geometry to
be a simple, low-order poloidal field, which supports our view
that the period is correct. Below we use this period in analyzing the
magnetic field morphology of WD\,1953-011.
At the same time it is important to note that another estimate of
the rotational period
(P~=~1.4418~days) obtained by \citet{BM05} from
differential
photometry of this MWD is similar to, but formally different from, our
result. Following the next section, where we
establish the magnetic field morphology of the low-field component,
we will discuss this difference in more detail.

\section{Modeling the weak-field component of the magnetic field of
WD\,1953-011}
\label{Model}

To verify that the behavior of the weak-field component of
WD\,1953-011 is consistent
with a nearly dipolar geometry, we have followed the schematic
method proposed by \citet{Lan97}. This method has already been
described and applied to establish the
magnetic field morphology of the weak field
white dwarf WD0009+501 (see \citet{VBMF05} for details).
For this reason here we do not explain all the modeling details, but restrict
ourselves to the presentation of the results.

In this paper we model the phase-resolved
measurements of the mean longitudinal field and mean field modulus of the
weak-field component within the framework of a pure dipole and
dipole+quadrupole field.
The phase-resolved observables for the weak-field component
which we use as input data are obtained by binning measurements in phase
and averaging. The binned data are presented in Table~\ref{tbl5}.

The dipole or dipole plus quadrupole models depend on
the following 10 parameters:
\begin{itemize}
\item[--] $B_\mathrm{d}$ and $B_\mathrm{q}$, the dipole and quadrupole
	  strength, respectively;
\item[--] $v_\mathrm{e}$, the stellar equatorial velocity;
\item[--] $i$, the inclination of the stellar rotation axis to the line
	  of sight;
\item[--] $\beta$, the angle between the dipolar axis and
	  the rotation axis;
\item[--] $\beta_1$ and $\beta_2$, the analogues of $\beta$ for the
	  directions identified by the quadrupole;
\item[--] $\gamma_1$ and $\gamma_2$, the azimuthal angles of the unit
	  vectors of the quadrupole;
\item[--] $f_0$, the ``reference'' rotational phase of the model;
\item[--] $v_\mathrm{e} sin i$, the projected stellar rotation velocity.
\end{itemize}
The angles $i$, $\beta$, $\beta_1$, $\beta_2$ range from $0\degr$ to
$180\degr$, while $\gamma_1$, $\gamma_2$, $f_0$ range from $0\degr$ to
$360\degr$. The rotational period $P = 1.448$ and the limb-darkening
constant
$u$, which also affects the expressions of the magnetic observables,
are taken as fixed. (Note that the pure dipole model would retain as
free parameters only $B_\mathrm{d}$, $v_\mathrm{e}$, $i$, $\beta$,
and $f_0$.)

For the stellar mass, \citet{BRG95}
gave the value of $0.844\,M_\odot$, which together with known
surface gravity of WD\,1953-011 ($log\, g = 8.412$, \citet{BRG95})
correspond to a stellar radius of about $0.0095\,R_\odot$.
This parameter and the period were then used
to estimate the equatorial and projected velocities of the star.

For the limb-darkening coefficient, we adopted the value of $u=0.5$.
Note that, as discussed by \citet{Bag00}, the results
of the modeling are only slightly influenced by the $u$ value.
The best-fit parameters are:\\

{\it A) Dipole }\\
%%%%%%%%%%%%%%%%%%%%%%%%%%%%ARRAY1%%%%%%%%%%%%%%%%%%%%%%%%%%%%%%%%%%%%%%%
\begin{displaymath}
\begin{array}{lcrcl}
%\hline
	       i  &=&  14\degr  &\pm&  10\degr \\
	    \beta &=&  14\degr  &\pm&  10\degr \\
	      f_0 &\approx& 352\degr  &   &    \\
	B_{\rm d} &=& 108       &\pm&  5\,kG \\
	v_{\rm e} &=& 0.33      &\pm&  0.05\,km\,s^{-1} \\
	v_{\rm e} sin i &=& 0.08&\pm&  0.03\,km\,s^{-1} \\
%\hline
\end{array}
\end{displaymath}
%%%%%%%%%%%%%%%%%%%%%%%%%%%%%%%%%%%%%%%%%%%%%%%%%%%%%%%%%%%%%%%%%%%\\
\vspace*{0.5cm}

{\it B) Dipole + quadrupole}\\

%%%%%%%%%%%%%%%%%%%%%%%%%%%%ARRAY2%%%%%%%%%%%%%%%%%%%%%%%%%%%%%%%%%%%%%%%
\begin{displaymath}
\begin{array}{lcrcl}
%\hline
	       i  &=&  18\degr  &\pm&  10\degr \\
	    \beta &=&   8\degr  &\pm&  10\degr \\
	      f_0 &\approx& 357  &&    \\
	  \beta_1 &=&  22\degr  &\pm&  10\degr \\
	  \beta_2 &=&  24\degr  &\pm&  10\degr \\
	 \gamma_1 &\approx&  77\degr  &&   \\
	 \gamma_2 &\approx& 243\degr  &&   \\
	B_{\rm d} &=& 178       &\pm&  30\,kG \\
	B_{\rm q} &=& 233       &\pm&  30\,kG \\
	v_{\rm e} &=& 0.33      &\pm&   0.05\,km\,s^{-1} \\
	v_{\rm e} sin i &=& 0.1      &\pm&   0.05\,km\,s^{-1} \\
%\hline
\end{array}
\end{displaymath}
%%%%%%%%%%%%%%%%%%%%%%%%%%%%%%%%%%%%%%%%%%%%%%%%%%%%%%%%%%%%%%%%%%%

Note that, as explained by \citet{Bag00}, the available
observations do not allow one to distinguish between two magnetic
configurations symmetrical about the plane containing the rotation
axis and the dipole axis. Such configurations are characterized by the
same values of $B_\mathrm{d}$, $B_\mathrm{q}$, $v_\mathrm{e}$,
$\gamma_1$, $\gamma_2$, $f_0$, while the remaining angles are related by
%%%%%%%%%%%%%%%%%%%%%%%%%%%%ARRAY%%%%%%%%%%%%%%%%%%%%%%%%%%%%%%%%%%%%%%%%
\[
\begin{array}{lccccr}

 (&i,          &\beta,          &\beta_1,          &\beta_2          &)
 \phantom{\;.}                                                \\[0.1cm]
 (&180\degr -i,&180\degr -\beta,&180\degr -\beta_1,&180\degr -\beta_2&)\;.
\end{array}
\]
%%%%%%%%%%%%%%%%%%%%%%%%%%%%%%%%%%%%%%%%%%%%%%%%%%%%%%%%%%%%%%%%%%%%%%%%%

Due to the fact that the spin axis angle is close
to a pole-on orientation and due to the small number of available
observables, the error bars on the derived quantities are fairly large.
For the same reasons, there are some uncertainties in the
determinations of all the parameters considered together. However,
despite these weakness, we are able to obtain some conclusions about the
most probable geometry of the white dwarf's global field.

The best fit of the dipole+quadrupole model applied to the observations is
shown by solid lines in Fig.~\ref{fig9}. For comparison, the dashed line
shows the fit obtained using the pure dipolar morphology. As is evident,
the dipolar model does not reproduce the observations well.
Examination of the reduced $\chi_r^2$ statistics shows that the quality of
the dipole+quadrupole fit ($\chi_r^2=0.22, 0.43$ for longitudinal field
and field modulus, respectively)
is significantly better than the pure dipole fit ($\chi_r^2=1.9, 2.2$).
We therefore conclude that the large-scale weak-field component of
WD\,1953-011 is better modeled by the superposition of a
dipole and quadrupole components.

\section{High-field component}
\label{Active}

\subsection{Migrating magnetic flux tube?}

In contrast to the well-organized, nearly sinusoidal variation of the
weak-field component, the phase behavior of the high-field structure
exhibits a number of peculiar features that made it impossible to model
these data as a simple low-order multipole:

\begin{itemize}

\item
According to the
measurements of the Zeeman-split satellite spectral features in the
H$\alpha$ wings, the mean field modulus of the strong-field component
does not show any noticeable variation during the star's rotation.
(Due to rotation, we see variation of the flux intensities from the
strong-field area, but the corresponding Zeeman displacement is nearly
constant.) The most likely explanation \citep{MFMW00}
is that there is an area with a nearly uniformly distributed
strong magnetic field. This explanation, if true, suggests that
the strong-field component has a localized geometry and cannot
be understood as a high-field term in the multipolar expansion
of the star's general field. Averaging all the data we determine
$<|B_{spot}|> = 515 \pm 7$~kG.

\item
The Zeeman pattern attributed to the strong-field component becomes visible
at rotational phases $\phi = 0.25 - 0.7$ and demonstrates variation in the
flux intensities that suggests rotational variability of the projected effective
size $S$ of the magnetic spot. The projected area of the strong-field
structure varies from zero to about 12\% of the
disk, consistent with the study of \citet{MFMW00}.
This observable can be used as an additional
parameter to test the rotational period of the star.
However, using this quantity to search for the period we did
not find a regular signal at any period within the
tested 5-year time base. Moreover, phasing the data with
the magnetic ephemeris characterized by the rotational period of
1.448 days, the resultant phase curve of the spot size variation (Fig.\,10)
appears distorted in comparison to the well-organized
behavior of the weak-field component phased with the same ephemeris.
We observe a small relative phase shift between the data obtained
with different telescopes that
may indicate a possible secular longitudinal drift
of the strong-field component.

\item
The averaged longitudinal field
of the strong field area varies from zero (when the area is invisible)
to about 450~kG (Table~\ref{tbl4}) which is comparable to the averaged
mean field modulus ($\approx 515$\,kG) of the field. This suggests a
deviation of the strong field component from any of low-order multipolar
geometries for which the difference between the full vector and
its longitudinal projection should much larger
(for example, for a centered dipole field the difference should be at least
2.5 times, \citet{S50}).
\end{itemize}

The last point suggests the presence of an essentially
vertical orientation of the magnetic field lines relative to
the star's surface, typical for local magnetic flux tubes in cool,
convective stars (the Sun for example). If the
geometry is a tube seen as a local magnetic
spot in the photosphere, we may also expect the above-mentioned secular
drift. To our knowledge and by an analogy to the
Sun, such fields are expected to show dynamical activity like migration
over the star's surface and be associated with dark spots that might
produce photometric variability of the star. Significant photometric
variability of WD\,1953-011 has been established
\citep{Wad03,BM05}.
However, in this paper we are unable to establish the association of the
darkness and magnetic spots for reasons which we discuss below.

\subsection{Geometry and location of the magnetic spot}

Measuring the longitudinal magnetic field of the strong-field component
as described in Sect.\,5 we noted that the estimates of the
longitudinal field of the strong-field area might be
affected by our simplifying assumptions. In order to control
these measurements we have also directly modeled the observed polarization and
intensity H$\alpha$ spectra of the star without parametrizing the surface field
components. In addition to the analysis of the Stokes $I$ and $V$
spectra, the observed $Q$ and $U$ spectra were also taken into consideration.
As can be seen in Fig.\,1, linear polarization Stokes $Q,U$ signatures are
detected only marginally at a few rotational phases. However,
this information can also be used to constrain the magnetic geometry of the
degenerate. Note that different magnetic geometries may produce similar
Stokes $V$ spectral features due to axial symmetry of circular polarization
provided by the longitudinal projection of the field.
Linear polarization restricts the strength and orientation of the
transverse field that, together with circular polarization, makes it
possible to resolve the geometry spatially. The observed $Q$ and $U$
spectra are mainly noise, but we may try to use them in terms of the upper limits.

In our model we examined several
low order multipolar magnetic field geometries, integrating over the surface elementary
(taken at a single surface element) Stokes $I,V,Q$, and $U$ spectra calculated
for various field strengths and
orientations of the magnetic field lines. The technique we used
to calculate the elementary spectra deserves some special explanation.
\\

{\it a) Simulation of the Stokes $I,V,Q$ and $U$ synthetic H$\alpha$ spectra.}
\\

Generally, accurate simulation of the split Balmer profiles in spectra of
strong magnetic
white dwarfs requires detailed computations of the main opacity sources
under the influence of strong magnetic fields. To our knowledge, these
computations have not yet been tabulated for practical use. For this reason,
a self-consistent solution of the transfer equation for the line profiles in
spectra of strong-magnetic white dwarfs cannot be performed without
special consideration of
additional parameters (related, for example, to the Stark
broadening in the presence of a strong magnetic field, \citet{J92}).
However, in case of the weak-field degenerates we may restrict
ourself to a zero-field solution similar to that presented by
\citet{WM79} or by \citet{SBJ92}. The method assumes that if
the Stark broadening dominates the line opacity, the total opacity
can be calculated as the sum of individual Stark-broadened Zeeman components.
The Stark broadening is suggested to be taken as ``non-magnetic'' in this
case. Under these simplifying assumptions we may simulate the local,
elementary Zeeman spectra using one of the following two ways:

\begin{itemize}
\item[i] to compute the transfer equation for all Stokes parameters
at given strength and orientation of a local magnetic field line
calculating the H$\alpha$ opacities as described, or (alternatively)
\item[ii]
to select a ``template'' H$\alpha$ profile typical for
a zero-field white dwarf with the same pressure-temperature
conditions as in WD\,1953-011 for construction of the
elementary Zeeman spectra. (In other words, we may try to construct from this template
profile individual Zeeman $\pi-$ and $\sigma-$ components, parametrizing
their magnetic displacement and relative intensities, and additively combine
them to obtain the elementary $I,V,Q,U$ H$\alpha$ profiles.)
\end{itemize}

The first, direct method of atmospheric calculations for WD\,1953-011
requires special theoretical tools which are outside the scope of
this observational paper.
The second, simplified method, which we will use, seems to be rather rough
due to the fact that the
fluxes from the individual $\pi-$ and $\sigma-$ components obtained
by using the zero-field template profile are generally not additive
(whereas their corresponding opacities can be added in the transfer equation).
Nevertheless, in the linear guess approximation they can be taken
as additive and the method can also be applied.
Besides, testing this method on some standard, well-studied
magnetic Ap/Bp stars we have obtained satisfactory results modeling their
observed polarizations. This allowed us to conclude that
the method is reasonably accurate.

Thus, to calculate the elementary $I,V,Q,U$ H$\alpha$ profiles
we adopted the use of the zero-field H$\alpha$ template profile
which was constructed from the observed I-spectra obtained at those moments,
when the strong-field Zeeman pattern is not seen. The Stark parts of the profile
were obtained by averaging the $I$ profiles at the rotation phases 0 and
0.91 (see Fig.\,1) in which the strong-field features are not seen. The
central ``zero-field'' Doppler profile
was adopted to reproduce in the model procedure the observed
low-field magnetic broadening of the H$\alpha$ cores at phases 0 and 0.91.
The necessary individual profiles of the non-displaced $\pi-$ and displaced
$\sigma-$ components were obtained by entering the normal Zeeman displacements
according to the orientation of the local magnetic field: the circularly polarized
$\sigma-$ components are displaced according to the longitudinal
projection of the local magnetic field, and the linearly polarized
$\sigma-$ components are displaced by the transverse field.

In order to model the polarization H$\alpha$ profiles, the
relative intensities of the central $\pi-$ and displaced $\sigma \pm$
components were computed from the prjection of the local field vector onto
the plane on the sky and on the line of sight
as described by \citet{U56} (for a qualitative explanation see also
\citet{L80}). The final intensities of the
$\pi-$ and $\sigma-$ components were
obtained by renormalisation such that
the total sum of the fluxes from
all the components be equal to the flux from the zero-field template
H$\alpha$ profile.

Finally, the elementary $\pi-$ and $\sigma-$ components were combined
to construct from them
the elementary $I,V,Q,U$ spectra by simulation of the ordinary
and extraordinary beams given by a polarimetric analyzer.
For example, simulating the H$\alpha$ Stokes $V$ profile,
all the components except the circularly polarized $\sigma-$ components
are equally distributed between the beams. The circularly polarized
$\sigma_+$ components are absent in one of the beams, and the oppositely
polarized $\sigma_-$ components are absent in the other beam. The final
Stokes $V$ H$\alpha$ profile was obtained by subtraction of the ordinary
from extraordinary beams and devision of the result by the total flux.
The $Q$ and $U$ spectra were obtained in a similar way: the linearly polarized
$\sigma$ components and central $\pi$ component are distributed
between the beams according to the projection of the local magnetic field
onto the plane of the sky.

After the determination of the elementary $I,V,Q,U$ spectra given by
a magnetic geometry in all surface elements, we finally integrated and
averaged them over the disc. For the limb-darkening coefficient
we adopted the value of $u=0.5$.\\

{\it b) Modeling the field geometry in WD\,1953-011 by simulation of
the observed polarization spectra.}

In Sect.\,5) we have concluded, that estimates of the longitudinal magnetic
field based on the weak-field approximation are accurate enough to make it
possible to model the geometry of the low-field component separately from the
strong-field component in the manner as demostrated in Sect.\,7. For this
reason, and in order to reduce a number of variables, we use those results
(obtained in Sect.\,7) as input and non-changeable parameters in the tomography.
We just note, that modeling
the observed spectra obtained at those time moments, where the spot is not
seen, we have confirmed, that
the observations (Stokes-V spectra) can be better fit by the
dipole+quadrupole geometry of the low-field component with parameters performed in Sect.\,7 (
case {\it B}).
Examination of the pure dipole model (case {\it A}) gives no satisfactory results
and we do not use this case here.

Modeling the strong-field area as an additional
harmonic in the low-order (lower than octupole) multipolar expansion
we were unable to reproduce the observations.
The fit does not provide the necessary contrast in the observed
Zeeman patterns at those phases where the polarization and intensity spectra
demonstrate the weak- and strong-field Zeeman features together.
However, assuming the strong-field component to be concentrated
into a localized area having maximum projected size of about 12 per
cent of the disk, the strong-field Zeeman spectral features can be
well-reproduced with an average magnetic field of $550\pm 50$\,kG.
Practically the same result has been obtained by
\citet{MFMW00}.

To model the strong-field area we tested two simplified localized geometries:
a ``contrast spot'' with a homogeneously distributed, essentially
vertical magnetic field, and a ``sagittal'' geometry with a
strong vertically-oriented central magnetic field, that smoothly decreased
to zero at the spot edges. Generally, both geometries are able to describe
the Stokes $I$ and $V$ spectra with more or less acceptable accuracy.
The first model, however, does not provide us with a good fit of the
$Q$ and $U$ spectra due to the presence of the sharp (and non-physical) jump
of the field intensity at the edges of the strong-field area. For this
reason we do not discuss this case in detail.

The ``sagittal'' geometry of the strong-field area was constructed
by using a modified model of a centered dipole: about 45\% of the spot's area (central parts)
have the dipolar distribution with polar field $Bp = +810$\,kG at the center.
The remaining 55\% of the external dipolar field is artificially modulated
to have a gradual decrease to zero at the edges of the area. This model provides
a good fit of the Stokes $I,V$ spectra and reasonable reproduction of
the linear polarization Q-,U-spectra, as shown in Fig.\,11 where we also
illustrate the tomographic portrait of the white dwarf's magnetosphere.

Despite the fact that we obtain such a good agreement of the ``sagittal'' geometry
with the observables from the strong-field area, we do not claim that
this geometry is fully correct in all details (for example, the model does not
control conservation of magnetic flux). Similar to the case of the
weak-field component, the most natural way to study
the strong-field area is to describe it as a strong-field
feature resulting from the superposition of several high-order
harmonics in the multipolar expansion. At this time we are unable
to study this term using any combination of the first hamonics higher than
octupole, but we do not exclude that the use of the highest terms
of different polarities and intensities will resolve the problem.

However, this result clearly demonstrates a qualitative difference in the
morphologies of the strong-field area and the global field of the white dwarf.
From the model we establish with a very high probability that the
strong-field area has a localized structure with essentially vertical
orientation of the magnetic field lines. The physical size of the
area is about 20\% of the star's surface giving maximum 12\% projection
on the disc. The spot is located at an angle of about $67^\circ$ with respect
to the spin axis, providing a maximum longitudinal field strength
of about 400\,kG. These results are in good agreement with the
measurements of the longitudinal field  of the spot.

\section{Discussion}
\label{Discuss}

We have presented new low-resolution spectropolarimetric observations
of the magnetic white dwarf WD\,1953-011. From these
observations and observations of previous authors we have determined
the star's rotation
period, mean longitudinal field, mean
field modulus, and surface field morphology.
Let us finally summarize these results.

\begin{itemize}
\item[1)] Our present picture of WD\,1953-011 consists of a MWD with
relatively smooth,
low-field global magnetic field component, and a high-field
magnetic area.
\item[2)]
The low-field component demonstrates regular periodicity with
period $P~=~1.4480 \pm 0.0001$~days.
We interpret this as the rotational period
of the white dwarf. The long-term stability
of the surface and longitudinal magnetic fields of the low-field component
enable us to interpret this component as a fossil poloidal
magnetic field consisting of dipolar and quadrupolar harmonics with
the following basic parameters:\\

1) the inclination of stellar rotation axis
	     $i = 18\degr \pm  10\degr$; \\
2) the angle between the dipolar axis and the rotation axis
$\beta = 8\degr \pm  10\degr$; \\
3) the dipole strength $B_{\rm d} = 178 \pm 30$\,kG; \\
4) the quadrupole strength $B_{\rm q} = 233 \pm  30$\,kG. \\

\item[3)]
The strong-field component exhibits a peculiar localized structure.
The mean field modulus of the spot
$|B_{spot}|$ is estimated to be $515 \pm 7$~kG, which
is consistent with the results presented by \citet{MFMW00}.
The longitudinal magnetic field of the spot varies with
rotational phase from $< 300$~kG to about $400$~kG.
Comparing the mean field modulus with the maximum longitudinal field
we suggest that the geometry of the high-field spot may be
similar to a magnetic flux tube with vertically-oriented magnetic field
lines. The spot is located at an angle of
$\approx 67\degr$ with respect to the spin axis.
\end{itemize}

Our results suggest that the magnetic field
of WD\,1953-011 consists of two physically different morphologies -
the fossil poloidal field and an apparently induced magnetic spot. To our knowledge,
fossil, slowly decaying global magnetic fields are organized in a nearly
force-free poloidal configuration \footnote{According to the basic
properties of the Maxwell stress tensor (see, for instance,
\citet{PA79}) the magnetic field ${\bf B}$ creates in the atmospheric plasma
an isotropic pressure  $\frac{B^2}{8 \pi}$ and tension $\frac{B_i B_j}{4 \pi}$
directed along the magnetic lines of force. While neighboring lines of force
of a magnetic field try to expand due to the magnetic pressure, tension tends
to compensate for this effect. The force-free configuration is possible only
when the gradient of the magnetic pressure is fully compensated by the
tension forces.}.
In contrast, if the suggested vertical orientation of the magnetic
field lines in the spot is correct, the uncompensated ``magnetic
pressure''${^3}$ of such a localized field may dominate against the tension
causing a strong impact on the pressure-temperature balance in the photosphere
of the degenerate. This may produce a temperature difference between the
strong-field area and other parts of the star's surface. As a result we
may expect rotationally-modulated photometric variability of WD\,1953-011
(as observed by \citet{Wad03} and \citet{BM05}).
For these reasons (and by analogy to sunspots) such
fields might be unstable if not supported by other
dynamical processes such as differential rotation, and
may therefore exhibit secular drift with respect to the stellar rotation axis.

In the above context we note that significant photometric variability of
WD\,1953-011 has been established \citep{Wad03,BM05}. Also, remarkably,
the authors had encountered
problems analyzing the periodicity of the variable differential flux.
When individual epochs of their photometric data are phased
according to the rotation period of about 1.45 days, the resultant folded
lightcurves are smooth and approximately sinusoidal. However,
they had difficulty obtaining an acceptable fit to all epochs
of photometric data considered simultaneously \citep{Wad03}.
Besides, periodograms \citep{BM05} obtained separately
for their 7 individual observing
runs indicate {\it a significant spread in the period
distribution}. The individual peaks are stochastically distributed
around a rotation period $P \approx 1^d.45$ from $P \approx 1^d.415$
to $P \approx 1^d.48$, also suggesting a probable phase shift from epoch
to epoch with characteristic times from tens to hundred of days.
Combining all the data, they establish their version ot the rotation period
$P = 1.^d441769(8)$\,days. This period is significantly different from
the period derived by us from the behavior of the global field of the
star (P~=~1.448~$\pm$~0.0001~days).

Indirectly, these facts suggest a
physical relationship between the darkness and magnetic spots, and their
possible secular migration. Unfortunately, our spectroscopy and
the available photometry were obtained at different epochs, making it impossible
to study this relationship in this paper. Examination of these
problems will be among the goals of our further study of this magnetic
degenerate upon carrying out the necessary simultaneous photometric
and spectral observations of this star.

\begin{acknowledgements}
Our thanks to L.~Ferrario, P.~Maxted, and C.~Brinkworth for providing details of
individual spectroscopic and photometric measurements of WD\,1953-011.
We are also especially grateful to Stefan Jordan
(our referee) for valuable comments,
suggestions and his high estimation of our work.
GV is grateful to the Korean
MOST (Ministry of Science and Technology, grant M1-022-00-0005) and KOFST
(Korean Federation of Science and Technology Societies) for providing
him an opportunity to work at KAO through the Brain Pool program.
GAW and JDL acknowledge Discovery Grant support from the Natural Sciences and
Engineering Research Council of Canada. This study was also partially
supported by KFICST (grant 07-179). Based on observations collected at the
European Southern Observatory, Chile (ESO program 67.D-0306(A))
\end{acknowledgements}

\newpage
%%%%%%%%%%%%%%%%%%%%%%%%%%%%%%%%TABLE1%%%%%%%%%%%%%%%%%%%%%%%%%%%%%%%%%%%
\begin{table}
\caption{\label{tbl1} Spectral and spectropolarimetric observations
of WD\,1953-011: column~1 is the Julian Date of the midpoint of the
observation, column~2 is the exposure time, and column~3 reports
the telescope used in the observations ({\bf AAT} indicates the high
resolution spectroscopy presented by \citet{MFMW00}).
For data obtained with the VLT
the exposure times are presented for the three consecutive
$I,V$ / $Q$ / $U$ modes of observations (in this case the midpoint
corresponds to observations of the Stokes $I,V$ parameters).
}
\begin{tabular}{|l|c|c|l|}
\hline
\hline
JD  & $Exp$ (sec)& Stokes&Telescope\\
\hline
2450676.955&     600 &$I$& AAT  \\
2451391.948&     600 &$I$& AAT   \\
2451391.955&     600 &$I$& AAT   \\
2451391.962&     600 &$I$& AAT   \\
2451392.059&     1800&$I$& AAT   \\
2451392.957&     1800&$I$& AAT   \\
2451393.066&     1800&$I$& AAT   \\
2451393.106&     1800&$I$& AAT   \\
2451393.943&     1200&$I$& AAT   \\
2451393.958&     1200&$I$& AAT   \\
2451393.973&     1200&$I$& AAT   \\
2451393.988&     1200&$I$& AAT   \\
2451394.003&     1200&$I$& AAT   \\
2452048.801  &   840/1200/1200  &$I,V/Q/U$&  VLT\\
2452048.893  &   840/1200/1200  &$I,V/Q/U$&  VLT\\
2452076.671  &   840/1200/1200  &$I,V/Q/U$&  VLT\\
2452076.883  &   840/1200/1200  &$I,V/Q/U$&  VLT\\
2452078.722  &   840/1200/1200  &$I,V/Q/U$&  VLT\\
2452078.879  &   840/1200/1200  &$I,V/Q/U$&  VLT\\
2452079.672  &   840/1200/1200  &$I,V/Q/U$&  VLT\\
2452079.892  &   840/1200/1200  &$I,V/Q/U$&  VLT\\
2452087.621  &   840/1200/1200  &$I,V/Q/U$&  VLT\\
2452087.670  &   840/1200/1200  &$I,V/Q/U$&  VLT\\
2452087.722  &   840/1200/1200  &$I,V/Q/U$&  VLT\\
2452087.768  &   840/1200/1200  &$I,V/Q/U$&  VLT\\
2452505.290  &   3600  &$I,V$&      BTA\\
2452505.327  &   3600  &$I,V$&      BTA\\
2452505.360  &   3600  &$I,V$&      BTA\\
2452505.397  &   3600  &$I,V$&      BTA\\
\hline
\hline
\end{tabular}
\end{table}
%%%%%%%%%%%%%%%%%%%%%%%%%%%%%%%%TABLE1%%%%%%%%%%%%%%%%%%%%%%%%%%%%%%%%%%%
\newpage

%%%%%%%%%%%%%%%%%%%%%%%%%%Table2%%%%%%%%%%%%%%%%%%%%%%%%%%%%%%%%%%%%%%
\begin{table}
\caption{\label{tbl2} Determinations of the mean modulus $|B_G|$
of the weak-field component. Column\,1 is the Julian Date, Col.\,2
and Col.\,3 are the equivalent widths of the H$\alpha$ core ($EW_{core}$)
and associated error bar, Col.\,4 and Col.\,5 are the inferred
field strength $|B_G|$ and its error bar $\sigma$\,(kG), Col.\,6
is the telescope used. Uncertainties at the measured equivalent
widths are calculated as a noise fraction of the flux (due to Poisson noise)
in the total flux under the line profile.
The mean field modulus and its uncertainty obtained from the high-resolution
spectroscopy with the {\bf AAT} are taken from \citet{MFMW00}.
Uncertainties at the calibrated field strengths (observations with the
{\bf VLT} and {\bf BTA}) result from regression errors in the
$EW_{core}$\,--\,$|B_G|$ relationship shown in Fig.\,4.
}
\begin{tabular}{|l|c|c|c|c|l|}
\hline
\hline
JD  & $EW_{core}$& $\sigma$  & $|B_G|$ $ (kG)$& $\sigma (kG)$ &Telescope\\
\hline
2450676.955  & 1.040   & 0.008  & 91  & 5  &  AAT\\
2451391.948  & 1.073   & 0.008  & 93  & 4  &  AAT\\
2451391.955  & 1.105   & 0.016  & 100 & 4  &  AAT\\
2451391.962  & 1.045   & 0.008  & 93  & 4  &  AAT\\
2451392.059  & 1.068   & 0.012  & 93  & 2  &  AAT\\
2451392.957  & 0.905   & 0.012  & 83  & 1  &  AAT\\
2451393.066  & 0.943   & 0.012  & 80  & 2  &  AAT\\
2451393.106  & 0.933   & 0.012  & 83  & 1  &  AAT\\
2451393.947  & 1.008   & 0.020  & 92  & 3  &  AAT\\
2451393.958  & 1.013   & 0.016  & 87  & 2  &  AAT\\
2451393.973  & 0.945   & 0.016  & 84  & 2  &  AAT\\
2451393.988  & 0.935   & 0.020  & 84  & 2  &  AAT\\
2451394.003  & 0.981   & 0.020  & 83  & 2  &  AAT\\
2452048.801  & 0.901   & 0.012  & 80  & 3  &  VLT\\
2452048.893  & 0.917   & 0.012  & 81  & 3  &  VLT\\
2452076.671  & 1.012   & 0.012  & 89  & 3  &  VLT\\
2452076.883  & 1.057   & 0.016  & 93  & 3  &  VLT\\
2452078.722  & 1.013   & 0.016  & 89  & 3  &  VLT\\
2452078.879  & 0.939   & 0.012  & 83  & 3  &  VLT\\
2452079.672  & 1.052   & 0.016  & 93  & 3  &  VLT\\
2452079.892  & 1.102   & 0.012  & 97  & 3  &  VLT\\
2452087.621  & 0.952   & 0.016  & 82  & 3  &  VLT\\
2452087.670  & 0.939   & 0.016  & 83  & 3  &  VLT\\
2452087.722  & 0.911   & 0.016  & 78  & 3  &  VLT\\
2452087.768  & 0.921   & 0.016  & 79  & 3  &  VLT\\
2452505.290  & 0.978   & 0.020  & 84  & 4  &  BTA\\
2452505.327  & 1.000   & 0.020  & 88  & 4  &  BTA\\
2452505.360  & 1.028   & 0.020  & 91  & 4  &  BTA\\
2452505.397  & 1.034   & 0.020  & 91  & 4  &  BTA\\
\hline
\hline
\end{tabular}
\end{table}
%%%%%%%%%%%%%%%%%%%%%%%%%%Table2%%%%%%%%%%%%%%%%%%%%%%%%%%%%%%%%%%%%%%
\newpage

%%%%%%%%%%%%%%%%%%%%%%%%%%%Table3%%%%%%%%%%%%%%%%%%%%%%%%%%%%%%%%%%%%%%
\begin{table}
\caption{\label{tbl3} Determinations of the mean modulus $|B_{spot}|$
of the strong-field component. Column\,1 is the Julian Date, Col.\,2
and Col.\,3 are relative area of the spot on the disc $S$ (in percent
of the full disc area)
and associated error bar obtained as a noise fraction of the flux in the
total flux under the strong-field satellite features. Col.\,4 and Col.\,5 are the
magnetic field strength $|B_{spot}|$ and its error bar $\sigma$\,(kG) obtained
as uncertainty in the determination of the satellite positions deblended by
Gaussians. Col.\,6 is the telescope used.)}
\begin{tabular}{|l|c|c|c|c|l|}
\hline
\hline
JD  & $S $(\%)& $\sigma$ (\%) & $|B_{spot}|$ $(kG)$& $\sigma (kG)$ &Telescope\\
\hline
2450676.955  & 13.1   & 0.6  & 521         & 40 &  AAT\\
2451391.948  &  9.5   & 0.6  & 513         & 30 &  AAT\\
2451391.955  & 12.0   & 1.2  & 495         & 30 &  AAT\\
2451391.962  & 11.7   & 0.6  & 494         & 30 &  AAT\\
2451392.059  & 12.0   & 0.6  & 527         & 15 &  AAT\\
2451392.957  & 0.9    & 0.9  &invisible    &    &  AAT\\
2451393.066  & 0.6    & 0.9  &invisible    &    &  AAT\\
2451393.106  & 0.7    & 0.9  &invisible    &    &  AAT\\
2451393.947  & 5.4    & 1.5  &invisible    &    &  AAT\\
2451393.958  & 2.4    & 1.2  &invisible    &    &  AAT\\
2451393.973  & 0.6    & 1.2  &invisible    &    &  AAT\\
2451393.988  & 0.7    & 1.5  &invisible    &    &  AAT\\
2451394.003  & 0.7    & 1.5  &invisible    &    &  AAT\\
2452048.801  & 0.6    & 0.9  &invisible    &    &  VLT\\
2452048.893  & 0.6    & 0.9  &invisible    &    &  VLT\\
2452076.883  & 7.8    & 1.2  & 520         & 15 &  VLT\\
2452078.622  & 3.2    & 1.2  &invisible    &    &  VLT\\
2452078.879  & 0.5    & 1.2  &invisible    &    &  VLT\\
2452079.672  & 8.4    & 1.2  & 529         & 30 &  VLT\\
2452079.892  & 12.3   & 0.9  & 511         & 15 &  VLT\\
2452087.621  & 0.6    & 1.2  &invisible    &    &  VLT\\
2452087.670  & 0.6    & 1.2  &invisible    &    &  VLT\\
2452087.722  & 0.7    & 1.2  &invisible    &    &  VLT\\
2452087.768  & 0.6    & 1.2  &invisible    &    &  VLT\\
2452505.290  & 10.4   & 1.2  & 500         & 35 &  BTA\\
2452505.327  & 12.3   & 1.2  & 492         & 40 &  BTA\\
2452505.360  & 12.3   & 1.2  & 524         & 35 &  BTA\\
2452505.397  & 10.8   & 1.5  & 502         & 45 &  BTA\\
\hline
\hline
\end{tabular}
\end{table}
%%%%%%%%%%%%%%%%%%%%%%%%%%%Table3%%%%%%%%%%%%%%%%%%%%%%%%%%%%%%%%%%%%%%
\newpage

%%%%%%%%%%%%%%%%%%%%%%%%%%%%%Table4%%%%%%%%%%%%%%%%%%%%%%%%%%%%%%%%%%%%%%
\begin{table}
\caption{\label{tbl4} Determinations of the longitudinal magnetic field
of the low- and high-field components of WD\,1953-011.
Column\,1 is the Julian Date, Col.\,2
and Col.\,3 are the  longitudinal field of the low-field component
and associated error bar, Col.\,4 and Col.\,5 are the deduced
longitudinal magnetic field of the strong-field component and its error
bar (``no'' means ``below the detection level''),
Col.\,6 is the telescope used.)}
\begin{tabular}{|l|c|c|c|c|l|}
\hline
\hline
JD  & $B_G^l$ $(kG)$& $\sigma$ $(kG)$ & $B_l^s$ $(kG)$& $\sigma$ $(kG)$ &OBS\\
\hline
2452048.801  & -41.5    & 1.5  & no          &    &  VLT\\
2452048.893  & -39.6    & 1.6  & no          &    &  VLT\\
2452076.883  & -41.0    & 1.6  & 430         & 70 &  VLT\\
2452078.722  & -42.9    & 1.8  &             &    &  VLT\\
2452078.879  & -42.2    & 1.7  & no          &    &  VLT\\
2452079.672  & -41.9    & 1.6  & 360         & 60 &  VLT\\
2452079.892  & -46.8   & 1.7   & 460         & 60 &  VLT\\
2452087.621  & -41.5   & 1.6   & no          &    &  VLT\\
2452087.670  & -39.8   & 1.7   & no          &    &  VLT\\
2452087.722  & -40.1   & 1.7   & no          &    &  VLT\\
2452087.768  & -40.1   & 1.5   & no          &    &  VLT\\
2452505.290  & -46.2   & 2.3   & 440         & 80 &  BTA\\
2452505.327  & -44.8   & 2.3   & 450         & 80 &  BTA\\
2452505.360  & -45.0   & 2.5   & no          &   &  BTA\\
2452505.397  & -42.0   & 2.7   & no          &    &  BTA\\
\hline
\hline
\end{tabular}
\end{table}
%%%%%%%%%%%%%%%%%%%%%%%%%%%%%Table4%%%%%%%%%%%%%%%%%%%%%%%%%%%%%%%%%%%%%%
\newpage

%%%%%%%%%%%%%%%%%%%%%%%%%%%%%Table5%%%%%%%%%%%%%%%%%%%%%%%%%%%%%%%%%%%%%%
\begin{table}
\caption{\label{tbl5} Phase-resolved observable magnetic
quantities of the weak-field component of WD\,1953-011. The first
column is rotation phase $\phi$ obtained according to the magnetic ephemeris
characterized by the rotational period of 1.448 days found here;
the second and third columns are the mean field
modulus or ``surface magnetic field'' $|B_G|$ and associated error bar;
the fourth and fifth columns are
the longitudinal field $B_l^G$ and its error bar.
}
\hspace*{1.5cm}
\begin{tabular}{|l|c|c|c|l|}
\hline
\hline
$\phi $  & $|B_G|$ & $\sigma $ & $B_G^l$ & $\sigma$ \\
 & $(kG)$ & $(kG)$ & $(kG)$ & $(kG)$ \\
\hline
0.05  & 81.5  & 4  & -40.5   &1.3 \\
0.15  & 82    & 2  &         &    \\
0.25  & 89    & 3  &         &    \\
0.35  & 94    & 3  & -44     &1.5 \\
0.45  & 93    & 2  &         &    \\
0.55  & 97    & 3  & -47     &2   \\
0.65  & 89    & 3  &         &    \\
0.75  & 86    & 4  & -42.2   &2   \\
0.85  & 83    & 1  & -41     &1   \\
0.95  & 78    & 2  & -40     &1   \\
\hline
\hline
\end{tabular}
\end{table}

\newpage

\begin{figure}
\centering
\hspace*{-1.cm}
\includegraphics[width=6.8cm, height=9.0cm, angle=0]{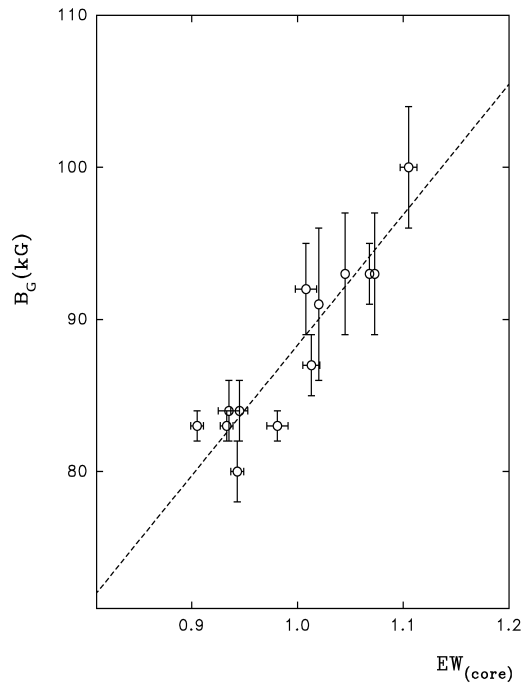}
\caption{
The relationship $EW_{core}$ -- $|B_G|$ obtained from the convolved
high-resolution spectra for which $|B_G|$ values are estimated by
\citet{MFMW00}. The dotted line is a linear fit of
the relationship.
}
\label{fig4}
\end{figure}

\begin{figure}
\centering
%\hspace*{-0.9cm}
\includegraphics[width=7.8cm, height=10.0cm, angle=270]{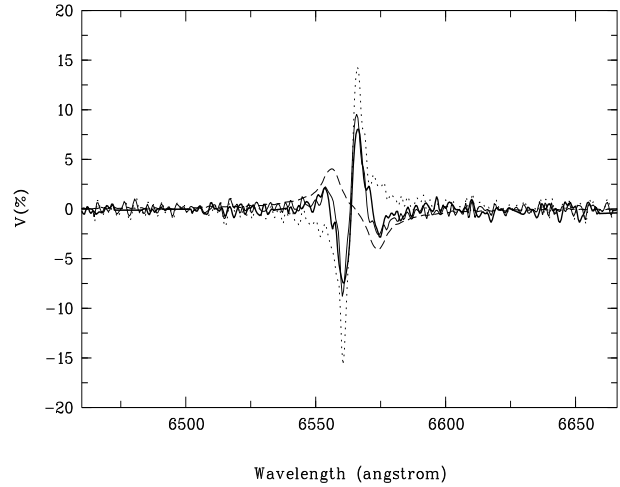}
\caption{
An example of the model technique as described in sect.~5. The thick
solid line is the Stokes $V$ observed spectrum containing strong circular
polarization from the strong-field component; the dotted and dashed lines are
modeled Stokes $V$ spectra of the weak- and strong-field components
respectively; the thin solid line is their sum.
}
\label{fig5}
\end{figure}

\begin{figure}
\centering
\includegraphics[width=8.5cm, height=5.5cm, angle=0]{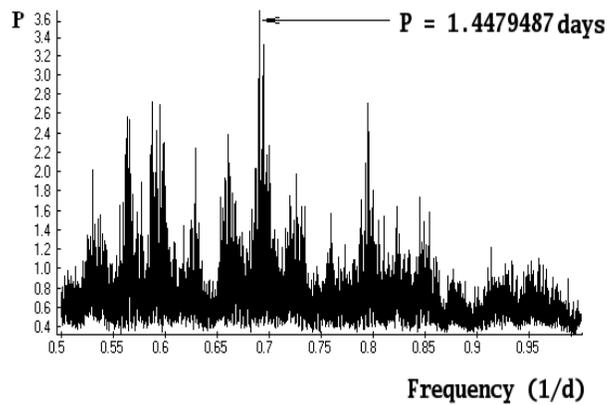}
\caption{Power spectrum of the magnetic field variations in WD\,1953-011.
}
\label{fig6}
\end{figure}

\begin{figure}
\vspace*{0.5cm}
\centering
\includegraphics[width=8.5cm, height=5.5cm, angle=0]{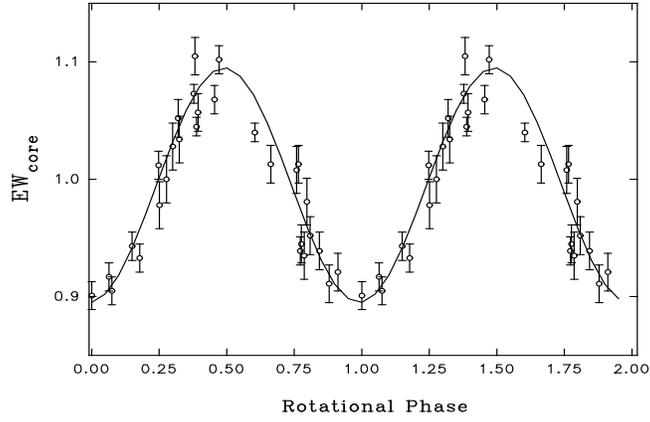}
\vspace*{0.5cm}
\caption{The phase curve of the equivalent widths at the H$\alpha$ core phased
with the 1.4480-day period.
}
\label{fig7}
\end{figure}

\begin{figure}
\vspace*{0.5cm}
\centering
\includegraphics[width=8.5cm, height=6.5cm, angle=0]{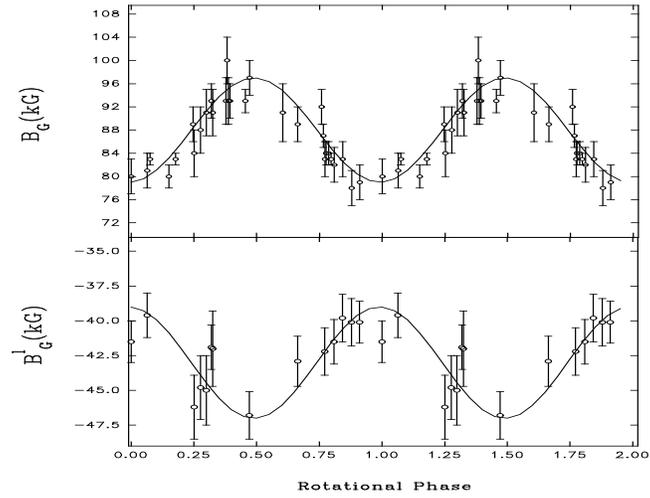}
\vspace*{0.1cm}
\caption{
The magnetic phase curves of WD\,1953-011 and their fits with the
1.448-day period. The upper plot illustrates variation of
the field modulus of the weak-field
component; the lower plot is variation of the its longitudinal magnetic field.
}
\label{fig8}
\end{figure}

\begin{figure}
\centering
\includegraphics[width=9cm, height=14cm, angle=0]{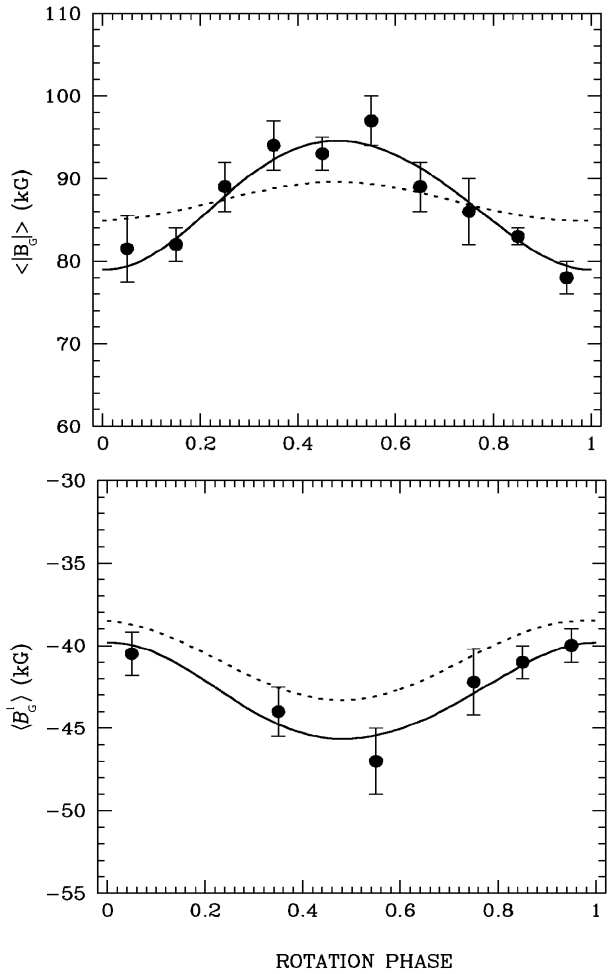}
%\vspace*{-3.1cm}
\caption{
Observations and modeling of mean longitudinal field (bottom panel) and
mean field modulus, or surface field (top panel). The dashed lines
show a fit obtained with dipole model. The solid lines show the best-fit
obtained by means of a dipole + quadrupole model.
}
\label{fig9}
\end{figure}

\begin{figure}
\centering
\includegraphics[width=12.5cm, height=8.5cm, angle=0]{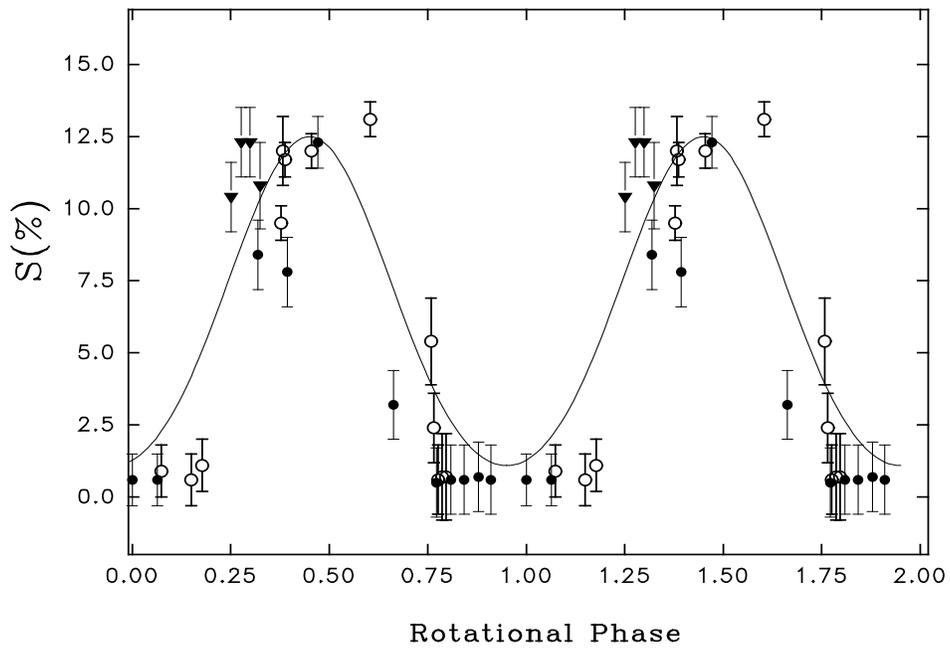}
%\vspace*{-3.1cm}
\caption{
Phase-resolved projection S of the strong-field area:
open circles illustrate the data from the AAT, filled circles
and triangles are the observations with the VLT and BTA respectively.
All the data have been phased according to the magnetic ephemeris obtained
in Sect.~6.
}
\label{fig10}
\end{figure}

\begin{figure*}
\centering
\includegraphics[width=13.5cm, height=16.0cm, angle=0]{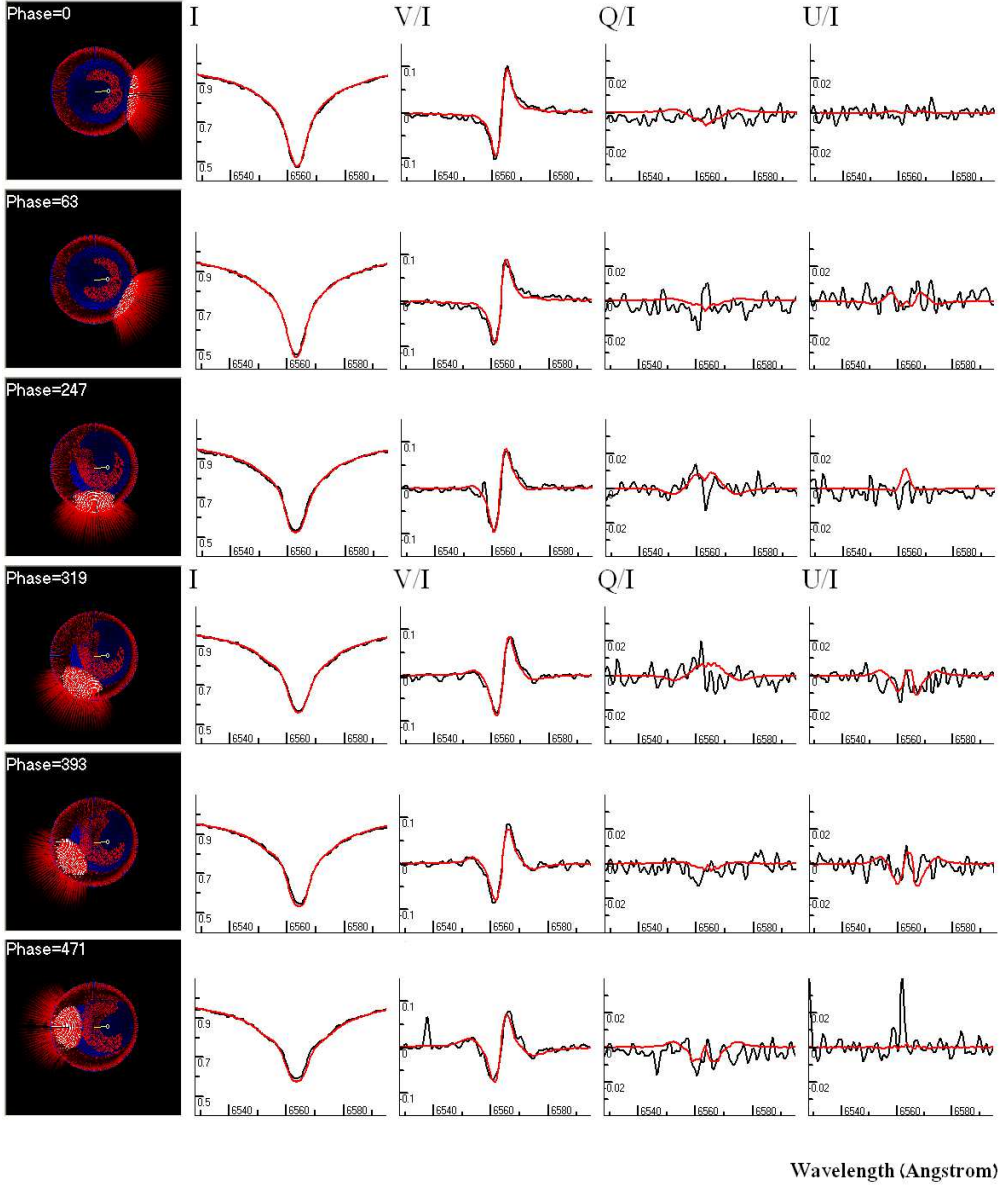}
%\vspace*{-3.1cm}
\caption{Model fits (red lines) of the observed (black lines)
Stokes $IVQU$ spectra obtained in observations with the VLT.
The corresponding tomographic portraits of the star's
magnetosphere and rotational phase (marked as {\bf Phase} ) are presented
at left. The strong-field area is shown by white. The magnetic field line of
force are red lines.
}

\label{fig11}
\end{figure*}
\newcounter{10}

\begin{figure*}
\centering
\includegraphics[width=13.5cm, height=16.0cm, angle=0]{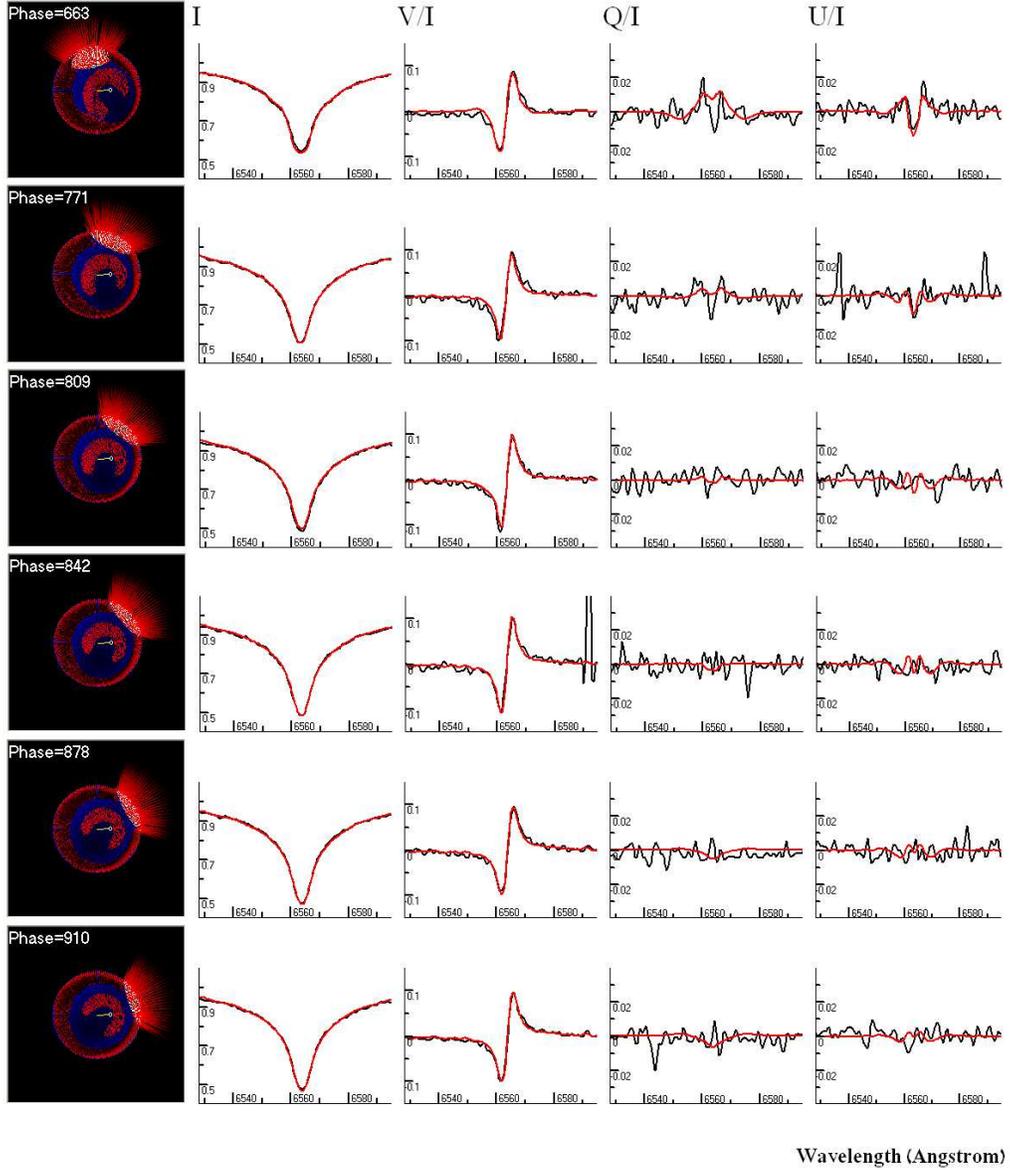}
%\vspace*{-3.1cm}
\caption{
continued.
}
\label{fig12}
\end{figure*}

\end{document}